\documentclass[12pt]{iopart}

\usepackage{iopams}
\usepackage{graphicx,amscd}
\usepackage{cite}

\begin{document}

\title{Particular spectral singularity in the continuum energies: a
  manifestation as resonances} \author{E. Hern\'andez$^{1}$,
  A. J\'auregui$^{2}$, D. Lohr$^{1}$ and A.  Mondrag\'on$^{1}$}
\address{$^{1}$Instituto de F\'{\i}sica, Universidad Nacional
  Aut\'onoma de M\'exico, Apdo. Postal 20-364, 01000 M\'exico
  D.F. M\'exico} \address{$^{2}$Departamento de F\'{\i}sica,
  Universidad de Sonora, Apdo. Postal 1626, Hermosillo, Sonora,
  M\'exico} \ead{mondra@fisica.unam.mx}

\begin{abstract}
We study the coalescence of two bound energy eigenstates embedded in
the continuous spectrum of a real Hamiltonian $H[4]$
and the singular point produced by this coalescence. At the
 singular point, the two unnormalized Jost eigenfunctions are no
longer linearly independent but coalesce to give rise to a bound state
eigenfunction embedded in the continuum. We disturb the potential
$V[4]$ by means of a truncation, this perturbation breaks the
singular point in two resonances.  The phase shift shows a jump of
magnitude $2\pi$ and the shape of the cross section shows two inverted
peaks, this behaviour is due to the interference between the two nearly
degenerate resonances and the background component of the Jost
function.
\end{abstract}


\maketitle

\section{Introduction}

In recent years the development and progress of quantum physics with
non-Hermitian operators has given rise to important accomplishments in
different fields
\cite{Moiseyev,QP-NHH,C-Bender,NHQM,narevicius,moiseyev0,berry,rotter0,brody},
and in particular in the study of the physics of exceptional points
\cite{QP-NHH,berry,Heiss0,Heiss1,Gunter0,Heiss3} and the spectral
singularities \cite{Gusein,Adrian0,Mosta,Bori,Mosta1,heiss0}.
%Exceptional points appear as a coalescence of both the eigenvalue and
%eigenfunction of a Hamiltonian of a given quantum system \cite{kato}.
%Exceptional points are often found in non-Hermitian Hamiltonian
%matrices of finite dimension
%\cite{QP-NHH,berry,Heiss1,Higgs,lefebvre1,cartarius,estrada} or in
%Hermitian Hamiltonians with non-selfadjoint boundary conditions
%\cite{ehs1,EH5}, in both cases they depend on a set of control
%parameters.  
Many theoretical
\cite{Heiss1,estrada,cartarius,lefebvre1,ehs1,EH5,latinne,Higgs,xai,
  Cavalli,Korsch,lefebvre,okolowicz,moiseyev,kalita} and
experimental \cite{bren1,Phil,D1,D2,Lee1,stehmann,Dietz,Atabek} works
are related to exceptional points produced by an accidental degeneracy
of resonant states.

Exceptional points in the real and continuous spectrum of a Hamiltonian
of infinite dimension \cite{Andrianov2,Sokolov2,longhi1} have
received much less attention than the case of non-Hermitian
Hamiltonians of finite dimension. These exceptional points are
associated with bound states embedded in the continuous energy of
scattering states.  Bound states embedded in the continuous energy were
first proposed by von Neumann and Wigner in 1929 \cite{vonN}. They
showed that certain spatially oscillating potentials could
support a bound state with energy above the potential barrier. Later
Stillinger et al \cite{stillinger1} proposed that these states could
be found in certain atomic and molecular systems, and in ultra-thin
layer structures of semiconductors. They showed \cite{stillinger2}
that lattices could be used to construct potentials that support bound
states with positive energies. 
The first experimental evidence of these states was reported by
F. Capasso et al \cite{FC1992} in a super lattice consisting of thin
superconducting layers of AlInAs/GaInAs.
J. Pappademos et al \cite{JP1993}
showed that, with methods of supersymmetric quantum mechanics, one can
construct potentials supporting bound states in the
continuum. A great number of examples of potentials that support a
bound state in the continuum have been studied \cite{MSP1982}.
%A. A. Stahlhofen \cite{Stahlhofen} constructed local
%potentials with positive eigenvalues for the one-dimensional
%Schr\"odinger equation. 
Bound states embedded in the
continuum have been recently observed in optical wave guide arrays
\cite{yonatan,steffen,corrielli}.
The first observation of a bound state associated with an exceptional
point (defect modes) in PT-symmetric optical lattice was done by
A. Regensburger et al \cite{regensburger}. 

A possible way to observe a bound state
embedded in the continuum is by perturbing the potential. T. A. Weber
and D. L. Pursey \cite{TAW1998} studied the $s-$wave scattering by a
von Neumann-Wigner type potential and showed that by truncating the
potential the bound state in the continuum manifest itself as a
resonance.   

%Some theoretical research
%about the formation of defect modes have been done in the framework of
%PT-symmetric optics
%\cite{longhi1,regensburger,zhou,molina,Xue}. S. Longhi, using a two
%times iterated discrete Darboux transformation \cite{longhi1}, showed
%that exceptional points in the continuum can appear in non-Hermitian
%optical lattices with engineered defects.

The Darboux transformation method is a powerful technique for the
generation of bound states in the continuum associated with
singularities of the function of the dispersion. These singularities
may be of diferent nature\cite{heiss0}, as poles of the function of the
  dispersion  or exceptional points of a
  non-Hermitian Hamiltonians\cite{longhi1}.

According to Kato \cite{kato} the exceptional points are characterized
by the coalescence of two eigenvalues and the corresponding
eigenfunctions.  Due to coalescence of their
eigenfunctions the scalar product is vanished.
These points are the beginning of branch points and
branch cuts on the surfaces of the energy as a function of the control
parameters of the system \cite{EH5,keck}.
However, in very few cases \cite{heiss0,EHS1} 
emphasis is placed on the important role of
the eigenvectors associated with these exceptional points.

A particular class of spectral singularity that has characteristics
similar to the exceptional points is that associated with zero energy
bound states. According to Newton \cite{Neuton} the problem of the
dispersion of a particle with a radial potential that admits bound
states may have resonances close to zero energy for $l > 0$. Newton
discusses the evolution of a resonant state through a zero energy
eigenstate to a bound state, by increasing the intensity of the
attractive potential. The resonance as well as the bound state are
poles of the dispersion function in the k-plane. It shows that a
resonance gives rise to two poles in lower plane k, which are
symmetrically located with respect to the imaginary axis k, by
increasing the intensity of the potential the two poles move towards
$k = 0$, and coalesce at this point and then continue moving along the
imaginary axis k in opposite directions see ref. \cite{Neuton}. 

In
2011 W. Heiss et al \cite{heiss0} reanalysed the single particle
dispersion around zero energy in relation to some recent experiments:
A Bose-Einstein condensate of neutral atoms with induced electromagnetic 
atractive interaction \cite{papado} in nanostructures \cite{mirosh} and
optics using continuous media with complex refractive index
\cite{guo,mark,mark2}. They studied the singular behaviour of the states 
of energy when the length of the interaction around a bound
state to zero energy is varied. The dispersion length $a_{\ell}$ is
defined by a expansion for energy eigenvalues \cite{Neuton,heiss0}
other than zero and has a first order pole when an eigenvalue at zero
energy is produced.  From an strength $v_0$ of the potential that
produces a bound state at $k = 0$ for orbital angular momentum $\ell >
0 $, Heiss \cite{heiss0} modifies the potential at $ v_0 + \epsilon $
and obtains new eigenvalues as an expansion in terms of the $\epsilon
$ potential intensity, $k_{1,2} = \sum_{n = 1}
C_{n}^{1,2}\sqrt{\epsilon^{n}}$, the square root in this expression is
a clear reminiscent of an exceptional point. They illustrate these
results for a square well of width $\pi $ and $\ell = 1$. Similarly,
they also find the expansion of the function of the dispersion around
$ k = 0$. However, when they analyse the eigenstates and the function
of the dispersion they find that the behaviour is different from that
of an exceptional point \cite{uva}.

In this paper we present an analytical and numerical study of a
particular type of spectral singularity of a Hamiltonian with a
potential $V{4}$ generated by four times iterated and completely
degenerate Darboux transformation\cite{Matveev}. 
This spectral singularity is 
reminiscent of the zero energy bound state of single particle
scattering for angular momentum greater than zero \cite{heiss0}.  The
potential $V[4]$ explicitly depends on two free parameters. Perturbing
the potential $V[4]$ we show that the particular spectral singularity
manifests itself as two resonances in the complex $k$-plane. Although
this potential is obtained without reference to any specific field
forces, it can be used to study some of the properties of Hamiltonian
operators.

This paper is organized as follows: In section 2 we generate a
Hamiltonian $H[4]$ by means of a four times iterated and completely
degenerated Darboux transformation. In section 3 we compute the Jost
solutions of $H[4]$ normalized to unit probability flux at infinity
and we show that, at $E = q^{2}$, the Wronskian of the two
unnormalized Jost solutions of $H[4]$ vanishes, this property
identifies this point as a particular spectral singularity in the
spectrum of the Hamiltonian $H[4]$.  In section 4 we perturb the
potential $V[4]$ with a cut off at a finite value $r = a$ and we show
that the particular spectral singularity manifest themselves as two
resonances.  In section 5 we study the interference between the two
resonances and the background term, which comes from an infinite
number of zeros of the Jost function.  A summary of the main results
and conclusions is given in section 6.

\section{The Hamiltonian $H[4]$}
A Hamiltonian that has some kind of spectral singularity in its real
and continuous spectrum may be generated by means of a four times
iterated and completely degenerated Darboux transformation
\cite{Matveev}.

The potential $V[4]$ obtained from the Darboux
transformation should not have any singularities that are not present in
the initial potential; this condition puts constraints on the 
potential $V[4]$  that fix the number of free
parameters as it will be shown in this section.

The radial Schr\"odinger equation is given by
\begin{equation}\label{1}
 \Bigl(-\frac{\partial^{2}}{\partial r^{2}} + V[4]\Bigr) \psi(r) = k^{2}\psi(r),
\end{equation}
with units $ 2m = 1$ and $\hbar = 1$.

The radial Schr\"odinger equation is defined in $[0, \infty )$ and we
will compute the regular solutions that satisfy the
boundary condition $\psi (0)= 0$.

According with Crum's generalization of the Darboux theorem
\cite{Crum}, the potential V[4] is obtained Wronskian
$W(\phi,\partial_{q}\phi,\partial^{2}_{q} \phi,\partial^{3}_{q}\phi)
\equiv W_{1}(q,r)$, and its derivatives with respect to $r$. The
transformation function, $\phi(q,r)$, is an eigenfunction of the free
particle radial Hamiltonian with eigenvalue $E =\ q^{2}$
\begin{equation}\label{3}
\phi(q,r) = \sin (qr + \delta(q)),
\end{equation}
and $\partial_{q}\phi$ is shorthand for $\partial \phi/\partial q$.
The phase shift $\delta(q)$ is a smooth function of the wave number
$q$.

The potential $V[4]$ is given by the equation 
\begin{equation}\label{4}
V[4] =-2\frac{1}{W^{2}_{1}(q,r)} \Bigl(W''_{1}(q,r)W_{1}(q,r) - W_{1}'^{2}(q,r)
\Bigr),
\end{equation}
the prime means differentiation with respect to $r$.

An explicit expression for $W_{1}(q,r)$ is the following
\begin{eqnarray}\label{5}
W_{1}(q,r) &=&   
16(q\gamma)^{4} -
12(q\gamma)^{2} + 8(q^{3}\gamma_{2})(q\gamma) - 12 (q^{2}\gamma_{1})^{2} \cr
&+& 24[(q^{2}\gamma_{1})(q\gamma) 
+ (q\gamma)^{2}]\cos 2\theta  
+ 3\sin^{2} 2\theta \cr
&+& [16(q\gamma)^{3} - 12q\gamma - 12q^{2}\gamma_{1} -
4q^{3}\gamma_{2}]\sin 2\theta,
\end{eqnarray}
where
\begin{eqnarray}\label{6}
\theta(r) &=& qr + \delta(q),  \gamma(r) =
\partial_{q}\theta = r + \gamma_{0}, \gamma_{0} = \partial_{q}\delta(q), \cr
\gamma_{1}&=& \partial^{2}_{q}\delta(q), \gamma_{2} = \partial^{3}_{q} 
\delta(q).
\end{eqnarray}
If the Wronskian $W_{1}(q,r)$ has a zero of order $n$ at the point
$r=r_{0}$ then the potential $V[4]$ has a centrifugal barrier of
the form $2n/(r - r_{0})^{2}$ and the potential diverges at that point.  
For large values of $r$, the dominant term in the right hand side
of eq.(\ref{5}) is $(q\gamma)^{4}$ which is positive and grows with
$r$ as $r^{4}$. Hence, for large values of $r$, the function $W_{1}(q,r)$ is 
positive and increasing. However, if the phase
$\delta(q)$ is left unconstrained, $W_{1}(q,r)$ could take a negative
value at the origin of the radial coordinate, $r = 0$, see figure 1,
in which case it should vanish for some positive value of $r$, giving
rise to a singularity of the potential $V[4]$ at that point.
%In figure 1 we show a Wronskian graph $W_{1}(q, r)$ for $\alpha, \ \beta
%$ in the neighborhood and at the singular point as a function of $r$.
\begin{figure}[ht]
\begin{center}
\includegraphics[width=245pt,height=150pt]{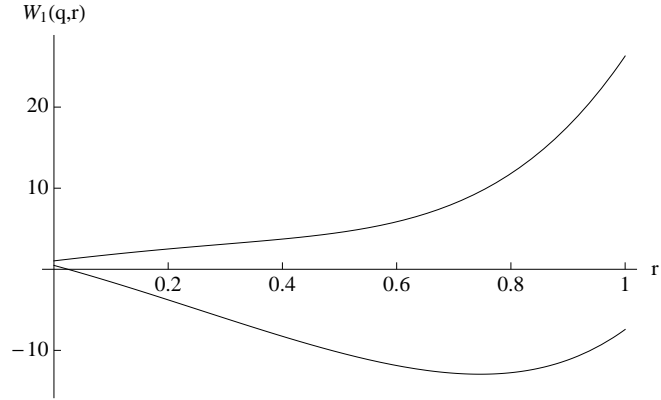}
\caption{The graph shows $W_{1}(q,r)$, as a function of $r$, for the
values of the parameters $\alpha = 1, q = 1 \ \mbox{and} \ \beta =-1$ and 
$5$. The function satisfies the condition $W_{1}(q,0) > 0$. For $\beta=-1$ 
the function takes negative values and the potential has divergences, for 
$\beta=5$ the function is positive and the potential is free of divergences.}
\end{center}
\end{figure}

In the case under consideration, the necessary condition for the
validity of the method of the Darboux transformation means that the
Wronskian $W_{1}(q,r)$ should not vanish for any positive value of
$r$. Therefore, to avoid the appearance of singularities in $V[4]$ we will set the condition
\begin{equation}\label{7}
W_{1}(q,0) > 0,
\end{equation}
which defines the phase shift $\delta(q)$, and from
equation (\ref{6}) the value of the functions
$\gamma_{0}$, $\gamma_{1}$, and $\gamma_{2}$ is computed.  An explicit
demonstration is given in Appendix A.

From equation (\ref{014}), we verify that the Wronskian $W_{1}(q,r)$, for
$r = 0$, is given by
\begin{equation}\label{8}
W_{1}(q,0) = \frac{12\beta^{2}}{(1 + (\alpha q - \beta)^{2})^2},
\end{equation}
where $\alpha$ and $\beta$ are two parameters that determine the
potential $V[4]$. In figure 1 we show the behaviour of the function
$W_{1}(q,r)$ for some values of $\alpha$, $q$ and $\beta$. For $\beta <0$ 
the function has an interval where it is negative, so it has points where
 is zero and the potential diverges in those points, whereas for $\beta> 0$ 
the function $W_{1}(q,r)$ is always positive. However, as
we will see later, these parameters are not independent they satisfy
the condition $\beta = 3\alpha q$.

\section{The Jost solutions in the 
spectrum of $H[4]$}

The two linearly independent unnormalized Jost solutions of $H[4]$,
that belong to the energy eigenvalues $E = k^{2}$ and behave as
outgoing and incoming waves for large values of $r$, are obtained from 
the Wronskians $W(\phi,...,\partial^{3}_{q}\phi, e^{\pm ikr})$ and
$W_{1}(q,r)$ \cite{Crum, Nicolas1}.

Notice that all terms in the last column of the Wronskian
$W(\phi,...,\partial^{3}_{q}\phi, e^{\pm ikr})$ are proportional to
$e^{\pm ikr}$. Hence, the unnormalized Jost solutions takes the form
\begin{equation}\label{9}
f^{\pm}(k,r) = \frac{W(\phi,...,\partial^{3}_{q}\phi,e^{\pm ikr})}{W_{1}(q,r)} =
\frac{1}{W_{1}(q,r)}w^{\pm}(k,r)e^{\pm ikr},
\end{equation}
where the function $w^{\pm}(k,r)$ is the reduced Wronskian
%\begin{equation}\label{10}
%w^{\pm}(k,r)e^{\pm ikr} =
%W(\phi,...,\partial^{3}_{q}\phi, e^{\pm ikr}).
%\end{equation}
 which is a complex function of its arguments
\begin{equation}\label{11}
w^{\pm}(k,r) = u(k,r) \pm iv(k,r).
\end{equation}
Explicit expressions for the functions $u(k,r)$ and $v(k,r)$ are given
by
\begin{eqnarray}\label{12}
u\left(k,r\right) &=& 16q^{4}\left(k^{2}-q^{2}\right)^{2}\gamma^{4}
-12q^{2}\left(k^{4}+6q^{2}k^{2}+q^{4}\right) \gamma ^{2} \cr && +
8\gamma_{2}q^{4} \left(k^{2}-q^{2}\right)^{2}\gamma
-12\gamma_{1}^{2}q^{4}\left(k^{2}-q^{2}\right)^{2} \cr &&+
24q^{2}\left[\left(k^{4}-4q^{2}k^{2}-q^{4}\right) \gamma^{2} + q\gamma
  _{1} \left(k^{4}-q^{4}\right) \gamma \right]\cos 2\theta \cr &&+
\left[16q^{3}\left(k^{4}-q^{4}\right) \gamma ^{3} -
  12q\left(k^{4}-4q^{2}k^{2}-q^{4}\right) \gamma \right. \cr
  &&-\left. 4\gamma _{2}q^{3}\left(k^{4}-q^{4}\right) - 12\gamma
  _{1}q^{2}\left( k^{4}-4q^{2}k^{2}-q^{4}\right) \right] \sin 2\theta
\cr &&+3\left( k^{4}+6q^{2}k^{2}+q^{4}\right)\sin ^{2}2\theta,
\end{eqnarray}
and
\begin{eqnarray}\label{13}
v(k,r) &=& 64 q^{4}k(k^{2} - q^{2})\gamma^{3} - 24q^{2}k(k^{2} + q^{2})\gamma 
+ 8\gamma_{2}q^{4}k(k^{2} - q^{2})  \cr
&&- 48\gamma_{1}q^{5}k + \bigl[32q^{4}k(k^{2} - q^{2})\gamma^{3} + 24q^{2}k(k^{2} 
+ q^{2})\gamma - 8\gamma_{2}q^{4}k \cr 
&&\times (k^{2} - q^{2}) 
+ 48\gamma_{1}q^{5}k\bigr]\cos 2\theta + [96q^5k\gamma^2  
- 48\gamma_{1}q^4k(k^2 - q^2)\gamma \cr
&&- 12qk(k^2 + q^2)]\sin 2\theta + 6qk(k^{2} + q^{2})\sin 4\theta.
\end{eqnarray}

For large values of $r$, the asymptotic behaviour of $w^{\pm}(k,r)$ is
dominated by the highest power of $r$. From eqs. (\ref{11}), (\ref{12})
and (\ref{13}) we get
\begin{eqnarray}\label{14}
w^{\pm}(k,r) = 16(k^{2} - q^{2})^{2}(qr)^{4}
 + O(r^{3}),
\end{eqnarray}
and from eq.(\ref{014}) it is clear that
\begin{equation}\label{15}
W_{1}(q,r) = 16(qr)^{4}\bigl[1 + O(r^{-1})\bigr],
\end{equation}
hence, for large values of $r$ the unnormalized Jost solutions are
given by
\begin{equation}\label{16}
f^{\pm}(k,r) = \bigl[(k^{2} - q^{2})^{2} + O(r^{-1})\bigr]e^{\pm ikr}.
\end{equation}

At infinity the Jost solutions are incoming and outgoing
waves, at the origin they have a finite value \cite{Neuton}.
The factor $(k^{2} - q^{2})^{2}$ is the flux of probability current at
infinity of the unnormalized Jost solutions.
Therefore, the Jost solutions of $H[4]$ normalized to unit
probability flux at infinity are
\begin{eqnarray}\label{17}
F^{\pm}(k,r) &=& \frac{f^{\pm}(k,r)}{(k^{2} - q^{2})^{2}} 
= \frac{1}{(k^{2} -  q^{2})^{2}}\frac{w^{\pm}(k,r)}{W_{1}(q,r)}
e^{\pm ikr}, \hspace{0.3 cm} k^{2} \ne q^{2}.
\end{eqnarray}

Each pair of linearly independent Jost solutions belongs to a point
$E_{k} = k^{2}$, with $k^{2} \neq q^{2}$, in the spectrum of $H[4]$.

The Wronskian of the unnormalized Jost solutions is readily computed
from (\ref{9}) and  eqs. (\ref{11}-\ref{13}) 
\begin{eqnarray}\label{18}
W(f^{+}(k,r), f^{-}(k,r)) = -2ik(k^{2} - q^{2})^{4}.
\end{eqnarray}
\begin{figure}[ht]
\begin{center}
\includegraphics[width=290pt,height=190pt]{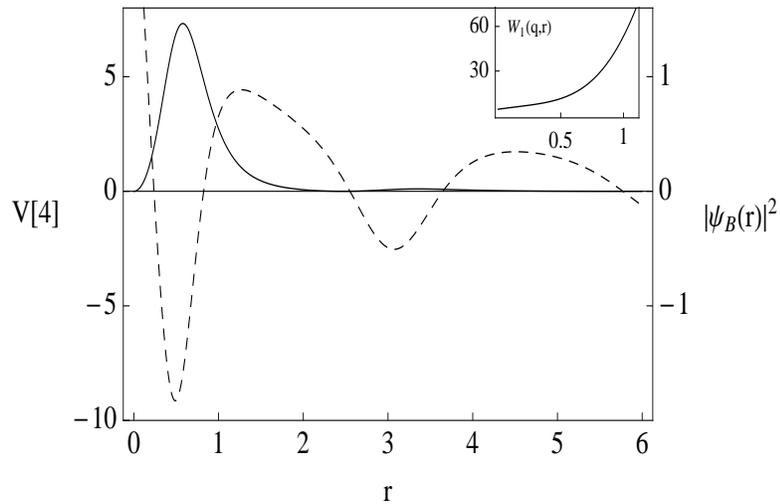}
\caption{The graph shows with a dashed line the potential $V[4]$ for
  values of the parameters $\alpha = 1, \beta = 3$ and $q =
  1$. The potential oscillates with an amplitude which decreases as
  $r^{-1}$ with increasing $r$, in the origin takes the value 19.55
  and the maximum of the first oscillation is at a height of 4.43. The
  probability amplitude of the normalized bound state solution as a
  function of $r$ is shown with a solid line. Note the difference of
  scale with the potential.  The bound state is confined in the first
  well of the potential. The inset illustrates the rapid growth and the
  absence of zeros of $W_{1}(q,r)$ for these parameters.}
\end{center}
\end{figure}
At the points $k^{2} = q^{2}$, the Wronskian of the two unnormalized Jost
solutions of $H[4]$ vanishes, then the two unnormalized Jost solutions
are no longer linearly independent and coalesce in the function 
\begin{eqnarray}\label{19}
f^{\pm}(q,r) &=& 4q^{2}\frac{24q^{2}}{W_{1}(q,r)}\bigl[-2q^{2}\gamma^{2}
\cos\theta + (q\gamma + q^{2}\gamma_{1})\sin\theta \cr
&+& \sin^{2}\theta\cos\theta\bigr]e^{\mp i\delta},
\end{eqnarray}
which is obtained from eq.(\ref{9}) when $k=\pm q$. From eq.(\ref{19}) it
follows the asymptotic behaviour of the function $f^{\pm}(q,r)$, which
goes to zero as $r^{-2}$, as $r\rightarrow \infty$. Hence,
$f^{\pm}(q,r)$ can be normalized and represents a bound state
eigenfunction embedded in the continuum and can be written as
\begin{eqnarray}\label{0010}
\psi_{B}(q,r) &=& \frac{24q^{2}}{W_{1}(q,r)}\bigl[-2q^{2}\gamma^{2}
  \cos\theta + (q\gamma + q^{2}\gamma_{1})\sin\theta +
  \sin^{2}\theta\cos\theta\bigr].
\end{eqnarray}
where $\gamma_{0},
\gamma_{1}$ and $\gamma_{2}$, as functions of $q$, are given explicitly in 
Appendix A. The bound state embedded in the continuum is zero at r = 0, 
this condition is guaranteed if the parameters satisfy the relation 
$\beta = 3\alpha q$. In this paper this condition is used with positive 
parameters.

The normalized probability amplitude of the bound state solution 
$\psi_{B}(q,r)$ and the potential $V[4]$, as functions of $r$, for the 
parameters $\alpha=1, q=1$ and $\beta=3$ are shown in figure 2. The bound 
state is confined in the first well of the oscillating potential for 
energy $E_{q} = q^{2}= 1$.

The Wronskian of the unnormalized Jost solutions, eq.(\ref{18}), goes
to zero as the fourth power of $(k-q)$ as a result of the coalescence
of four functions.  This property identifies the point $E = q^{2}$ as
a point associated with a particular spectral singularity in the
spectrum of the Hamiltonian $H[4]$.
This particular spectral singularity is associated with a double pole
in the normalization factor of the Jost eigenfunctions $F^{\pm}(k,r)$,
see equation (\ref{17}), but they are not associated with a double
pole in the scattering matrix. The scattering matrix and the cross
section are regular analytical functions of the wave number $k$ and
the Green Function has a first order pole \cite{Nicolas1}.  The square
root that characterizes an exceptional point is not present in this
kind of singularity.

\section{Truncated $V[4]$ potential}
The bound state in the continuum associated with the type of spectral
singularity presented in this paper is not related to the poles of the
scattering matrix and, therefore, is not possible to perform a direct
measurement. However, this state is formed in the first well of the
oscillating potential, its binding energy is $E_{q} = q^{2} = 1$, so
it is a fragile state and any disturbance can break the equilibrium
necessary for its formation, thus showing his presence in the complex
$k$-plane \cite{TAW1998}. The
advantage of this method of perturbation is that, although it may seem
artificial, it allows for an analytical study of the solutions.  In
this section we disturb the potential by cutting off its range at a
value $r = a$.  A cut off value of $a = 5000$ means that the
perturbation is very small; however, it is enough to show the breaking of
the bound state in the continuum into two resonances.

The radial Schr\"odinger equation is
\begin{equation}\label{21}
\left[-\frac{d^{2}}{dr^{2}} + V(r)\right]\varphi(k,r) = k^{2}\varphi(k,r),
\end{equation}
with the potential
\begin{equation}\label{22}
V(r) = \left\{
  \begin{array}{l l}
    V[4](r) & \quad r\leq a \\
    0 & \quad r>a
  \end{array} \right.
\end{equation}
$\varphi(k,r)$ is the regular wave function that is uniquely defined by the
boundary condition at $r=0$ \cite{Neuton}
\begin{equation}\label{23}
\lim_{r\rightarrow 0} r^{-1}\varphi(k,r) = 1.
\end{equation}
The  regular wave function  in both regions is given by the equation
 \cite{Neuton}
\begin{equation}\label{24}
\varphi (k,r) = \left\{\begin{array}{l l}
    \Phi (k,r) & \quad r\leq a \\
    \frac{i}{2k}\left[F(-k)e^{-ikr}-F(k)e^{ikr} 
\right] & \quad r>a
  \end{array} \right.
\end{equation}
in the potential region, $\varphi (k,r)$ is given by a linear
combination of the unnormalized Jost functions, eq.(\ref{9}). From the
boundary condition (\ref{23}) we get
\begin{eqnarray}\label{25}
\Phi(k,r)
&=& \frac{1}{h(k)}\frac{W_{1}(q,0)}{ W_{1}(q,r)}
\Bigl[ u(k,r)\Bigl(u(k,0)\sin kr - v(k,0)\cos kr\Bigr) \cr
&+& v(k,r)\Bigl(v(k,0)\sin kr + u(k,0)\cos kr \Bigr)\Bigr],
\end{eqnarray}
where
\begin{eqnarray}\label{251}
h(k)&=&u(k,0)\Bigl(\frac{\partial v(k,r)}{\partial r}\Bigr)_{r=0}-v(k,0)
\Bigl(\frac{\partial u(k,r)}{\partial r}\Bigr)_{r=0} \cr
&+& k (u^{2}(k,0)+v^{2}(k,0)).
\end{eqnarray}
At infinity, $\varphi(k,r)$ behaves as a linear combination of a free
incoming spherical wave plus a free outgoing spherical wave. In
eq.(\ref{24}), $F(-k)$ is the Jost function of the perturbed
problem \cite{Neuton}.

\subsection{Resonant state eigenfunctions}
To complete this study in this subsection we will give a brief
description of resonant state eigenfunctions or Gamow state
eigenfunctions.

Resonant states energy eigenfunctions $\psi_{n}(k_{n}, r)$ are
solutions of equation (\ref{21}), which vanishes at the origin,
\begin{eqnarray}\label{26}
\psi_{n} (k_{n}, 0 ) = 0,
\end{eqnarray}
and asymptotically behaves as purely outgoing waves,
\begin{equation}\label{27}
\lim_{r \rightarrow
  \infty}\Bigl[\frac{1}{\psi_{n}(k_{n},r)}\frac{d\psi_{n}(k_{n},r)}{dr}
  - ik_{n}\Bigr] = 0,
\end{equation}
that oscillate between envelopes that increase exponentially with
$r$, where  $k_{n}$ are the zeros of the Jost function, 
\begin{equation}\label{28}
F(-k_{n}) = 0,
\end{equation}
and the corresponding energy eigenvalues, $ E_{n}=k^{2}_{n}, $ are complex
with $Re E_{n} > Im E_{n}$.

The Gamow state eigenfunctions are given by \cite{Hernan}
\begin{equation}\label{29}
\psi_{n}(k_{n},r) =\frac{1}{N_{n}}\varphi(k_{n},r),
\end{equation}
where $N_{n}$ is the normalization constant written as \cite{berg}
\begin{equation}\label{30}
N_{n}^{2} = \frac{1}{i4k_{n}^{2}}F(k_{n})
\Bigl(\frac{dF(-k)}{dk}\Bigr)_{k=k_{n}}.
\end{equation}
The features described are illustrated in the following subsection,
where a graphic representation of the Gamow state eigenfunctions that
characterizes the resonances coming from the breakdown of the
exceptional point in the complex $k-$plane is shown.

\subsection{The phase shift and the cross section}
In this section we computed the phase shift $\delta(k)$ and the
  cross section $\sigma(k)$.

By matching the function $\varphi (k,r)$ and its derivative at the
boundary $r = a$ we get the following expression for the Jost function
\begin{equation}\label{31}
F(-k) = e^{ika}\left[\Phi'(k,a)-ik\Phi(k,a) \right]
\end{equation}
and
\begin{equation}\label{32}
F(k) = e^{-ika}\left[\Phi'(k,a)+ik\Phi(k,a)  \right],
\end{equation}
where $\Phi'(k,a)=\bigl(\partial\Phi(k,r)/\partial r \bigr)_{r=a}$.

From eq. (\ref{25}) the derivative of the regular 
solution is 
\begin{eqnarray}\label{33}
\Phi' (k,r) &=& \frac{1}{h(k)} \frac{W_{1}(q,0)}{W_1^{2}(q,r)} 
\bigl\{\bigl[u'(k,r)W_1(q,r)-u(k,r)W'_1(q,r) \cr
&-& kv(k,r)W_1(q,r) \bigr] (u(k,0)\sin kr - v(k,0)\cos kr) \cr 
&+&  \left[v'(k,r)W_1(q,r) - v(k,r)W'_1(q,r) + ku(k,r)W_1(q,r) \right] \cr
&\times&  ( v(k,0)\sin kr+u(k,0)\cos kr ) \bigr\} . 
\end{eqnarray}
\noindent
Substituting (\ref{25}) and (\ref{33}) in (\ref{31}) we get for the
Jost function the expression
\begin{equation}\label{34}
F(-k) = \frac{1}{h(k)}\frac{W_1(q,0)}{W_1^{2}(q,a)}e^{ika}
\left[d(k) + ig(k) \right]
\end{equation}
and for the function ${F}(k)$
\begin{equation}\label{35}
F(k) = \frac{1}{h(k)} \frac{W_1(q,0)}{
W_1^{2}(q,a)}e^{-ika}\left[d(k) - ig(k) \right],
\end{equation}
with
\begin{eqnarray}\label{36}
d(k) &=& \left[u'(k,a)W_1(q,a)-u(k,a)W'_1(q,a)-kv(k,a)W_1(q,a) \right] \cr
  &\times& (u(k,0)\sin ka -v(k,0)\cos ka) + 
\left[v'(k,a)W_1(q,a)-v(k,a) \right. \cr
&\times& \left. W'_1(q,a) + ku(k,a)W_1(q,a)\right](u(k,0)
\cos ka +v(k,0)\sin ka),
\end{eqnarray}
and
\begin{eqnarray}\label{37}
g(k) &=& -kW_1(q,a)\left[u(k,a)(u(k,0)\sin ka - v(k,0)\cos ka) \right. \cr 
&+& \left. v(k,a)(v(k,0)\sin ka + u(k,0)\cos ka)\right].  
\end{eqnarray}

The scattering matrix  $\mathbb{S}(k)$ is given by:
\begin{equation}\label{38}
\mathbb{S}(k) = \frac{F(k)}{F(-k)} .
\end{equation}

The zeros of the Jost function $F(-k)$ are the poles of
$\mathbb{S}(k)$. Substituting eqs. (\ref{34}) and
(\ref{35}) in (\ref{38}), we obtain
\begin{equation}\label{39}
\mathbb{S}(k) = \frac{e^{-ika}\left[d(k) - ig(k) \right]}
{e^{ika}\left[d(k) + ig(k) \right]}.
\end{equation}
\begin{figure}[ht]
\begin{center}
\includegraphics[width=260pt,height=150pt]{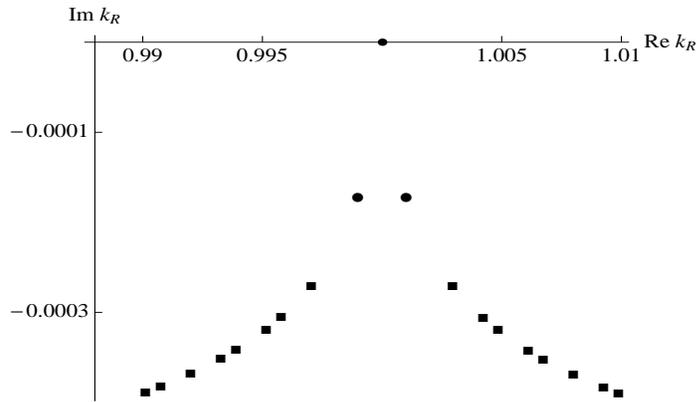}
\caption{The graph shows the zeros of the Jost function in the complex
  k-plane for the potential $V(r)$ described in the text with cut off
  parameter: $a=5000$. The mark on the real axis represents the point
  describing the particular spectral singularity, and the black points
  represent the resonances coming from the breakdown of this
  particular spectral singularity. The more distant zeros of
  the Jost function are represented by small squares.}
\end{center}
\end{figure}

From equation (\ref{39}) the poles of the $\mathbb{S}(k)$
matrix in the neighbourhood of the singular point $E_{q} = q^{2}$, are
obtained from the equation
\begin{eqnarray}\label{40}
 d(k)  + i g(k)  = 0. 
\end{eqnarray}

We solved numerically this equation for potential parameters $\alpha =
1$, $\beta = 3$ and the cut off parameter $a = 5000$, and found two
zeros of the Jost function in the fourth quadrant of the complex
$k-$plane close to $q = 1$, which correspond to the resonances in
which this particular spectral singularity is broken by disturbing the
system. These zeros are plotted in figure 3 with black points.  There
are other zeros of the Jost function more distant of the real axis of
the complex $k-$plane.  In figure 3 we show these zeros
with small squares.

As the cut off parameter increases, the two resonances plotted with
black points approach the real axis to coalesce at the point
representing this particular spectral singularity.
\begin{figure}[ht]
\begin{center}
\includegraphics[width=290pt,height=190pt]{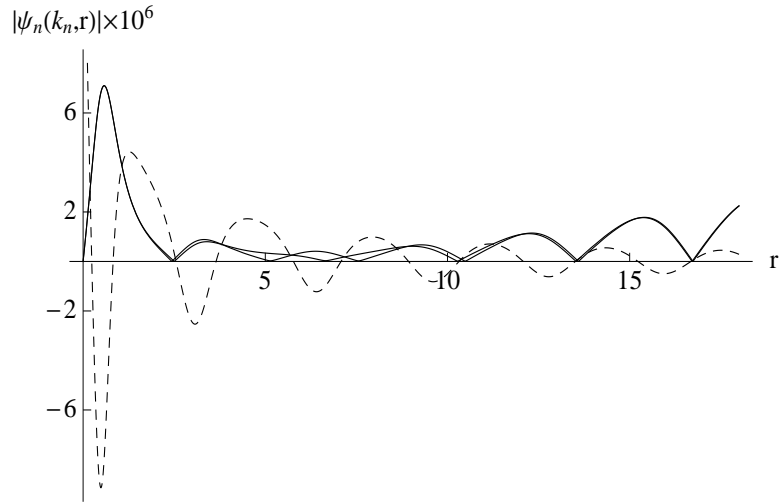}
\end{center}
\caption{The probability amplitudes of the normalized eigenfunctions
  of resonant states as a function of $r$ for the parameters $\alpha =
  1$, $\beta =3$ and the cut off $a=5000$ are shown with a continuous
  line. The perturbed potential $V(r)$ is shown by the dashed line.}
\label{graf2.2}
\end{figure}

In figure 4 we show, on the same scale of the potential $V[4]$, the
probability amplitudes of the normalized eigenfunctions of resonant
states corresponding to the resonances closest to the real axis $k$,
for the parameter values of the perturbed potential with $a = 5000$
and resonance wave numbers $k_{1} $ = 0.9989844032, $\Gamma_{1}/2$ =
0.0001730065 and $k_{2}$ = 1.0010155756, $\Gamma_{2}/2$ =
0.0001731296.  The resonances coming from the breaking of the point
representing the particular spectral singularity are formed in the
first well of the potential.  The resonances are nearly degenerate, as
we can see from the values of the real parts $k_{1}$ and $k_{2}$ of
wave numbers and their half-widths $\Gamma_{1}/2$ and $\Gamma_{2}/2$
and therefore the probability amplitudes of resonant states
eigenfunctions are almost indistinguishable.  In this figure the
tunneling of the probability amplitudes of resonant state
eigenfunctions through the oscillating potential is observed.

\begin{figure}[ht]
\begin{center}
\includegraphics[width=215pt,height=130pt]{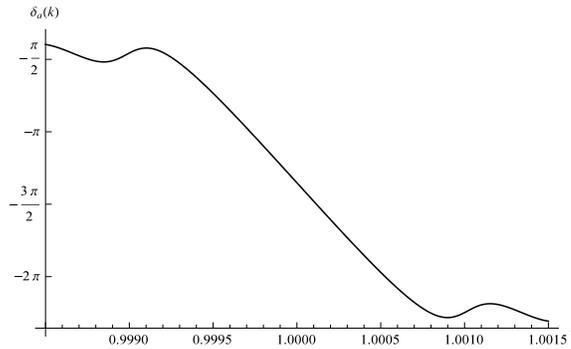}
\end{center}
\caption{Phase shift $\delta(k)$ as a function of the wave 
number $k$ for  $q=1$ and the parameters  $\alpha = 1$, $\beta
  =3$ and the cut off $a=5000$. The phase shift shows a jump of 
magnitude $2\pi$. }
\end{figure}

The scattering matrix $\mathbb{S}(k)$ given in eq. (\ref{39}) is
written as:
\begin{equation}\label{41}
\mathbb{S}(k) = e^{2i\delta_{a} (k)},
\end{equation}
where
\begin{equation}\label{42}
\delta_a (k) = -\arctan \frac{d(k)\sin ka + g(k)\cos
  ka}{d(k)\cos ka -g(k)\sin ka},
\end{equation}
is the phase shift of the perturbed potential. 

\begin{figure}[ht]
\begin{center}
\includegraphics[width=225pt,height=130pt]{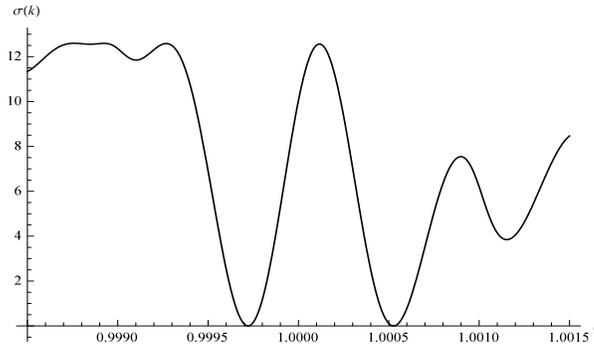}
\end{center}
\caption{Cross section $\sigma (k)$ as a function of the wave number
  $k$, for $q = 1$, and for the cut off $a = 5000$, and the values of
  the parameters $\alpha = 1$, $\beta =3$. The cross section shows two 
inverted peaks coinciding in position with the values of $\delta_{a}(k)=0$.}
\label{graf2.3}
\end{figure}

In figure 5 we show the phase shift $\delta_a (k)$ as a function of
the wave number $k$, and its behaviour in the neighbourhood of $k=1$
for the cut off parameter $a = 5000$, where it shows a jump of $2\pi$,
characteristic of the case where there are two nearly degenerate
resonances \cite{ehs1}. The jump starts near $-\pi/2$ and ends near 
$-5\pi/2$. In this case, as in any other potential that ends abruptly, 
the phase shift is influenced by the cut off through the exponential 
factor that appears in Jost function, see eq. (\ref{31}). This factor 
provides the term $-ka$ which dominates the increasing contribution of 
the phase shift in its passage through each resonance resulting in a 
negative phase shift, as in figure 5.

The cross section is defined as
\begin{equation}
\sigma (k) = \frac{4\pi}{k^2}\sin^2 \delta_a (k) .
\label{43}
\end{equation}
\noindent
Figure 6 shows the cross section as a function of the wave number
$k$. The inverted double peak is a feature of the two nearly
degenerate resonances.  Minima in the cross section are mainly
produced by interference effects in the first potential well and
tunneling through the oscillating potential, this effect is known as
Ramsauer-Townsend effect \cite{Neuton}.  Extremal points occur when the
phase shift passing through the values $-\pi$ and $-2\pi$, where
$\sigma(k)$ is a minimum; whereas the peak around $k =1.0001$ is due
to $\delta_{a}(k)$ passing through $-3\pi/2$, where $\sigma(k)$ is a
maximum.

\section{Interference of two close resonances}
In this section we will show the shape of the cross section, previously 
obtained, with the structure of two inverted peaks, as a result of the
interference of the two resonances closest to the real axis of the
complex $k-$plane and the background made up for distant resonances
and other no-resonant phenomena.

When the first and second absolute momenta of the potential exist, and
the potential decreases at infinity faster than any exponential or if
it vanishes identically beyond a finite radius, the Jost function
$F(-k)$ is an entire function of $k$ \cite{Neuton}. The entire function
of $k$, $F(-k)$, may be written in a form of an infinite product of
zeros according to Hadamard's form of the Weierstrass factorization
theorem \cite{hadamard} and, by using a theorem of Pfluger \cite{pfluger},
\begin{equation}\label{44}
F(-k) = F(0)\exp(ikR)\prod_{n = 1}^{\infty}\Bigl(1 - \frac{k}{k_{n}}\Bigr),
\end{equation}
where $R$ is the range of the potential, $F(0) = A k$ with $A$ a
constant and $\{k_{n}\}$ are the zeros of $F(-k)$ \cite{Neuton}.

In order to show the interference of the two resonances and the
background we write the Jost function $F(-k)$ in the explicit form of a
product of two zeros
\begin{equation}\label{45}
F(-k) = \left(k- k_1 + i \frac{\Gamma_1}{2} \right)\left(k- k_2 + i 
\frac{\Gamma_2}{2} \right)\exp(ika)D(k),
\end{equation}
the product $\exp (ika) D(k)$ to the background
component of the Jost function with
\begin{equation}\label{46}
D(k) = F(0)\frac{1}{(k_{1}- i\frac{\Gamma_{1}}{2})
(k_{2} -i\frac{\Gamma_{2}}{2})}\prod_{n = 3}^{\infty}\Bigl(1 -
\frac{k}{k_{n}}\Bigr).
\end{equation}
We consider a small perturbation of the potential by taking a large
value of the cut off parameter $a$. This guarantees that the doublet
of isolated resonances is close to the real axis of the complex
$k-$plane. The other zeros of $F(-k)$, contained in $D(k)$, are further
away from the real axis and also from the first pair of resonances and
correspond to distant resonances or other non-resonant phenomena,
thus the infinite product of zeros on the right hand side of equation
(\ref{46}) has a smooth behaviour of $k$, which can be written as
\begin{equation}\label{47} 
D(k) = \mu(k)(1 + i\lambda(k)),
\end{equation}
where $\mu(k)$ and $\lambda(k)$ are real and smooth functions of
$k$. From eqs.(\ref{45}), (\ref{46}) and (\ref{47}), the Jost function
takes the form
\begin{eqnarray}\label{48}
F(-k) &=& \mu(k)\{(\mathsf{Y}(k) - \lambda(k) \mathsf{Z}(k))\cos ka 
- (\lambda(k)\mathsf{Y}(k) + \mathsf{Z}(k))\sin ka  \cr
&+& i[(\mathsf{Y}(k) - \lambda(k) \mathsf{Z}(k))\sin ka 
+ (\lambda(k)\mathsf{Y}(k) + \mathsf{Z}(k))\cos ka]\},
\end{eqnarray}
where
\begin{eqnarray}\label{49}
\mathsf{Y}(k) &=& (k -k_{1})(k - k_{2})-\frac{\Gamma_{1}\Gamma_{2}}{4}, \\
\mathsf{Z}(k) &=& \frac{1}{2}[(k -k_{1})\Gamma_{2} + (k - k_{2})\Gamma_{1}].
\end{eqnarray}

The Jost function is written in terms of the phase shift as
\begin{equation}\label{51}
F(-k) = | F(k) | e^{-i\delta (k)},
\end{equation}

\begin{figure}[ht]
\begin{center}
\includegraphics[width=200pt,height=260pt]{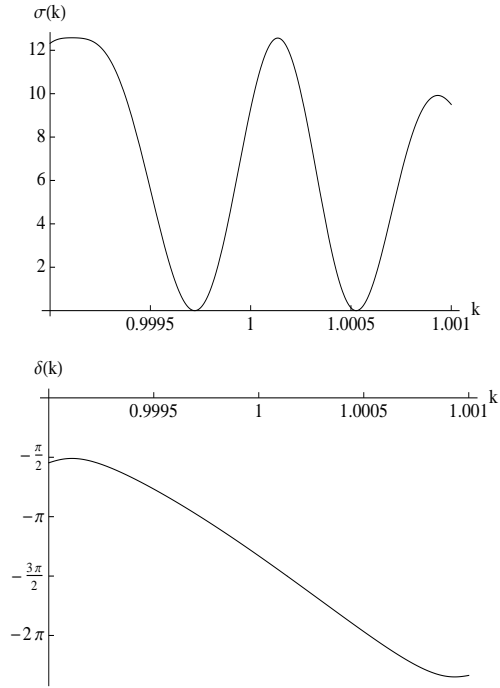}
\caption{The cross section $\sigma (k)$ and the phase shift $\delta
  (k)$ as function of $k$, calculated for the values of the resonances
  $k_{1} = 0.9989844032$, \ $\Gamma_{1}/2 = 0.0001730065$, $k_{2} =
  1.0010155756$, $\Gamma_{2}/2 = 0.0001731296$, and the parameters
  $\lambda_{0} = 1311.3931$, $\lambda_{1} = -1312.2167$ and a cut off $a =
  5000$.}
\end{center}
\end{figure}

and the  phase shift is given by 
\begin{eqnarray}\label{52}
\delta (k) = - \arctan \frac{(\mathsf{Y}(k)
  -\lambda(k)\mathsf{Z}(k))\sin ka + (\lambda(k)\mathsf{Y}(k) +
  \mathsf{Z}(k))\cos ka}{ (\mathsf{Y}(k) -
  \lambda(k)\mathsf{Z}(k))\cos ka - (\lambda(k)\mathsf{Y}(k) +
  \mathsf{Z}(k))\sin ka}.
\end{eqnarray}
The cross section $\sigma(k)$ is given by the expression
\begin{eqnarray}\label{53}
\sigma (k) = \frac{4\pi}{k^{2}(1 + \lambda^{2}(k))}
\frac{[(\mathsf{Y}(k) - \lambda(k)\mathsf{Z}(k))\sin ka +
    (\lambda(k)\mathsf{Y}(k) + \mathsf{Z}(k))\cos ka]^{2}}
{\mathsf{Y}^{2}(k) + \mathsf{Z}^{2}(k)}.\nonumber \\
\end{eqnarray}

From the  expressions for the phase shift and the cross
section we can see that the function $\lambda(k)$ gives a measure of
the interference between the two resonances and the background
term in the truncated potential $V(r)$.

For the computation of the cross section $\sigma(k)$ of the
approximation given by eq.(\ref{53}), we take the resonance wave numbers
obtained from the numerical calculation of the Jost function zeros
eq.(\ref{40}) for a cut off parameter $a = 5000$
\begin{eqnarray}\label{54}
k_{1} &=& 0.9989844032, \ \Gamma_{1}/2 = 0.0001730065 \\
k_{2} &=& 1.0010155756, \ \Gamma_{2}/2  = 0.0001731296
\end{eqnarray}
\begin{figure}[ht]
\begin{center}
\includegraphics[width=235pt,height=130pt]{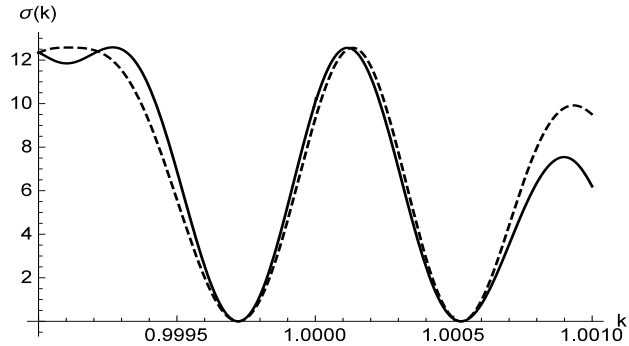}
\caption{The cross section $\sigma (k)$, as function of $k$. The
  continuous line is the numerically exact calculation, equation
  (\ref{43}), for the parameter values $\alpha = 1$, \ $ \beta = 3$,
  $q = 1$, and the cut off parameter $a = 5000$, the dashed line is the
  cross section of the approximation given by equation (\ref{53}),
  calculated for the values of the resonances $k_{1} = 0.9989844032$,
  \ $\Gamma_{1}/2 = 0.0001730065$, $k_{2} = 1.0010155756$, $\Gamma_{2}/2 =
  0.0001731296$, and the parameters $\lambda_{0} = 1311.3931$,
  $\lambda_{1} = -1312.2167$. }
\end{center}
\end{figure}
The assumption of the resonances being well isolated allows us to 
parametrise the function $\lambda(k)$  as a linear polynomial of $k$,
\begin{equation}\label{56}
\lambda(k) = \lambda_{0} + \lambda_{1} k,
\end{equation}
the parameter values $\lambda_{0} = 1311.3931$ and $\lambda_{1} =
-1312.2167$ are obtained from to fit the minima of the approximated
cross section to the minima of the exact cross section.

After the substitution of $\lambda (k)$ at eq. (\ref{52}), the phase
shift $\delta (k)$ in the neighbourhood of $k = 1$ exhibits a value
near $ -\pi/2$ for $k = 0.999$ and then the known jump of $2\pi$, as
can be seen from the graph for the phase shift in the lower part of
the figure 6. $2\pi$ jump of the phase shift is due to the
interference between the two nearly degenerate resonances and the
background parameterised through the $\lambda(k)$ function. In the
upper part of figure 7 we show the cross section as function of $k$,
it has the same resonant structure as the exact cross section.

Figure 8 shows the comparison of the results for the cross section, as
a function of the wave number $k$, obtained from the numerically exact
calculation with eq. (\ref{43}) and the results computed with the
approximation given by eq. (\ref{53}) in the neighbourhood of the
resonances.  The fit to the cross section using the approximation of
eq. (\ref{53}) is good enough in the sense that it reproduces the
shape of the resonant structure shown in the exact calculation and
allows us to give an explanation of this phenomenon.

\section{Summary and conclusions}
A study of a particular spectral singularity type in the continuous
spectrum of a real Hamiltonian $H[4]$ is presented
and discussed.  The Hamiltonian $H[4]$ and the free particle
Hamiltonian $H_{0}$ are isospectral. In the general case, to each
point in this continuous spectrum correspond two linearly independent
Jost solutions which behave at infinity as incoming and outgoing
waves. However, here we have shown that in the continuous spectrum of
$H[4]$ there is a point corresponding to a particular spectral
singularity type at $E_{q} = q^{2}$, this particular spectral
singularity is associated with a double pole in the normalization
factor of the Jost eigenfunctions normalized to unit flux at
infinity. At the singular point, the two unnormalized Jost
eigenfunctions are no longer linearly independent and coalesce to give
rise to a quadratically integrable bound state eigenfunction embedded
in the continuum. The bound state embedded in the continuum is formed
in the first well of the potential $V[4]$.  The perturbation of the
potential $V[4]$, with a cut off value $r = a$, manifests this
particular spectral singularity as two resonant states in the complex
$k-$plane. The two resonances are formed in the first well of the
perturbed potential. The phase shift shows a jump of magnitude $2\pi$
and the cross section shows two inverted peaks, where it vanishes, for
the values of $k$ where the phase shift is $ -\pi$ and $-2\pi$; and
it has a local maximum at the value of $k$ where the phase shift is
$-3\pi/2 $.  The shape of the cross section with two inverted peaks in
the neighbourhood of $k = q$ is due to the interference between the
two nearly degenerate resonances and the background component of the
Jost function. This phenomenon is known as the Ramsauer-Townsend
effect.

\section*{Acknowledgements}
We would like to thank Profs. E. Ley Koo and M. Mondrag\'on (IF-UNAM)
for some interesting discussions on the problem.  The authors are grateful to an anonymous reviewer for helpful comments. This work was
partially supported by CONACyT M\'exico under Contract No. 132059 and
DCEN-UNISON.


\appendix

\section{Computation of the Wronskian $W_{1}(k,r)$}

In this appendix we compute the explicit expression for the Wronskian
$W_{1}(q,r)$.
 
From eqs. (\ref{5}) and (\ref{6}) we have
\begin{eqnarray}\label{cero8}
W_{1}(q,0) &=& 16\Bigl(q\frac{d\delta(q)}{dq}\Bigr)^{4} -
12\Bigl(q\frac{d\delta(q)}{dq} \Bigr)^{2} +
8\Bigl(q^{3}\frac{d^{3}\delta(q)}{d q^{3}}\Bigr) 
 \Bigl(q\frac{d\delta(q)}{dq} \Bigr) \cr
&-& 12\Bigl(q^{2}\frac{d^{2}\delta(q)}{dq^{2}}\Bigr)^{2} +
24\Bigl[\Bigl(q^{2}\frac{d^{2}\delta(q)}{dq^{2}}\Bigr)
 \Bigl(q\frac{d\delta(q)}
  {dq}\Bigr) + \Bigl(q\frac{d\delta(q)}{dq}\Bigr)^{2}\Bigr]\cr
&\times& \cos2\delta(q)  +  3\sin^{2}2\delta(q) +  
\Bigl[16\Bigl(q\frac{d\delta(q)}{dq}\Bigr)^{3}-
  12\Bigl(q\frac{d\delta(q)}{dq}\Bigr) \cr
&-& 12\Bigl(q^{2}\frac{d^{2}\delta(q)}{dq^{2}}\Bigr) -
  4\Bigl(q^{3}\frac{d^{3}\delta(q)}{dq^{3}}\Bigr)\Bigr]  
\sin 2\delta(q).
\end{eqnarray}
 
To simplify the notation, we define a new function $t(q)$ as
\begin{equation}\label{cero9}
t(q):= \tan\delta(q),
\end{equation}
then
\begin{eqnarray}\label{cero13}
\sin 2\delta(q) = \frac{2t(q)}{1 + t^2(q)} \ \ \mbox{and} \ \ 
\cos 2\delta(q) = \frac{1 - t^{2}(q)}{1 + t^2(q)}.
\end{eqnarray}

Written in terms of $t(q)$, eq. (\ref{cero8}) takes the form
\begin{eqnarray}\label{cero10} 
Ẉ_{1}(q,0) &=& \frac{4
\Bigl(-t(q) + qt_{q}(q)\Bigr)}{\Bigl(1+t^{2}(q)\Bigr)^{2}}
\Bigl[3(-t(q)
+ qt_{q}(q)) + 6q^{2}t_{qq}(q) \cr 
&+& 2q^{3}t_{qqq}(q)\Bigr]  
- 3q^{4}\Bigl(\frac{d}{dq}(-t(q) + qt_{q}(q))\Bigr)^{2},  
\end{eqnarray}
in this expression $t_{q}$ is shorthand for $dt/dq$.

Now it is evident from (\ref{cero10}) that if $t(q)$ satisfies
\begin{equation}\label{cero11}
-t(q) + qt_{q}(q) = \beta, 
\end{equation}
the equation (\ref{cero10}) becomes an identity and the condition given in 
(\ref{7}) is satisfied provided that
\begin{equation}\label{cero15}
W_{1}(q,0) = \frac{12\beta^{2}}{(1 + t^2(q))^{2}}.
\end{equation}

Integrating (\ref{cero11}) we get
\begin{equation}\label{cero12-0}
t(q) = \alpha q - \beta,
\end{equation}
and according to equation (\ref{cero9}) the phase shift is given by
\begin{equation}\label{cero12}
\delta(q) =\arctan(\alpha q - \beta),
\end{equation}
in these expressions $\alpha$ and $\beta$ are free parameters, but
$\beta \neq 0$.

Once the phase shift $\delta(q)$ is known as an explicit function of $q$, the
functions $\gamma_{0}, \ \gamma_{1}$ and $\gamma_{2}$ are obtained
from its first, second and third derivative, respectively, 
\begin{eqnarray}\label{cero14}
\gamma_{0} &=&  \frac{\alpha}{1 + (\alpha q - \beta)^{2}}, \hspace{0.4cm} 
\gamma_{1} = -\frac{2\alpha^{2}(\alpha q - \beta)}
{(1 + (\alpha q - \beta)^{2})^{2}},  \cr
\gamma_{2} &=& - \frac{2\alpha^{3}(1 - 3(\alpha q - \beta)^{2})}
{(1 + (\alpha q - \beta)^{2})^{3}}.
\end{eqnarray}

With the help of these expressions and equation (\ref{5}) we get
for $W_{1}(q,r)$ the following expression
\begin{eqnarray}\label{014}
W_1(q,r) &=& \frac{12\beta^2}{(1+(\alpha q - \beta)^2)^2} +
\frac{24\beta \alpha q}{(1+(\alpha q - \beta)^2)^2} (\cos 2qr -1 ) \cr
 &+&  \frac{12\alpha q \left((\alpha q)^2 + \beta ^2 -
  1 \right)}{(1+(\alpha q - \beta)^2)^2} \sin 2qr + 16\bigl[(qr)^4 
+ \frac{4\alpha q}{(1+(\alpha q - \beta)^2)}   \cr
&\times&(qr)^3 + \frac{6(\alpha q)^2}{(1+(\alpha q - \beta)^2)^2}(qr)^2 
+ \frac{3(\alpha q)^3}{(1+(\alpha q - \beta)^2)^2} (qr)\bigr] \cr
&-& 12 \bigl[(qr)^2 + \frac{2\alpha q}{(1+(\alpha q -\beta)^2)}(qr)\bigr] 
+ 24 \bigl[(qr)^2  \cr
&+&  \frac{2\alpha q (1-\beta(\alpha q - \beta))}
{(1+(\alpha q - \beta)^2)^2}(qr)\bigr]  
\cos 2(qr + \delta(q)) +  \Bigl[16\bigl((qr)^3  \cr
&+&  \frac{3\alpha q}{(1+(\alpha q - \beta)^2)}(qr)^2 + 
\frac{3(\alpha q)^2}{(1+(\alpha q -  \beta)^2)^2} (qr)\bigr) -12(qr)\Bigr]  \cr
&\times& \sin 2(qr + \delta(q)) 
+  3 \Bigl[\frac{1-6(\alpha q-\beta)^2 + (\alpha q -
    \beta)^4}{(1+(\alpha q - \beta)^2)^2}  \cr
&\times&  \sin^2 2qr + \frac{4(\alpha
    q-\beta)(1-(\alpha q - \beta)^2)}{(1+(\alpha q - \beta)^2)^2}
\sin 2qr \cos 2qr \Bigr].
\end{eqnarray}

\end{document}